\documentclass[useAMS,usenatbib]{mnras}

\usepackage{natbib}
\usepackage{tabu,booktabs}
\usepackage{amsmath}
\usepackage{xcolor}
\usepackage[english]{babel}
\usepackage{graphicx}
\usepackage{xspace}

\usepackage{longtable}
\usepackage{threeparttablex}
\usepackage{array}

\newcommand{\hMpc}{{\ifmmode{h^{-1}{\rm Mpc}}\else{$h^{-1}$Mpc}\fi}}
\newcommand{\Mpc}{{\ifmmode{{\rm Mpc}}\else{Mpc}\fi}}
\newcommand{\hkpc}{{\ifmmode{h^{-1}{\rm kpc}}\else{$h^{-1}$kpc}\fi}}
\newcommand{\kpc}{{\ifmmode{ {\rm kpc} }\else{{\rm kpc}}\fi}}
\newcommand{\kms}{{\ifmmode{ {\rm km\,s^{-1}} }\else{ ${\rm km\,s^{-1}}$ }\fi}}
\newcommand{\hMsun}{{\ifmmode{h^{-1}{\rm {M_{\odot}}}}\else{$h^{-1}{\rm{M_{\odot}}}$}\fi}}
\newcommand{\Msun}{{\ifmmode{{\rm M}_{\odot}}\else{${\rm M}_{\odot}$}\fi}}
\newcommand{\Mhalo}{{\ifmmode{M_{\rm halo}}\else{$M_{\rm halo}$}\fi}}
\newcommand{\Rvir}{{\ifmmode{R_{\rm vir}}\else{$R_{\rm vir}$}\fi}}
\newcommand{\Mvir}{{\ifmmode{M_{\rm vir}}\else{$M_{\rm vir}$}\fi}}
\newcommand{\Mstar}{{\ifmmode{M_{\rm star}}\else{$M_{\rm star}$}\fi}}
\newcommand{\Vrot}{{\ifmmode{V_{\rm rot}}\else{$V_{\rm rot}$}\fi}}
\newcommand{\ltsima}{$\; \buildrel < \over \sim \;$}
\newcommand{\gtsima}{$\; \buildrel > \over \sim \;$}
\newcommand{\lsim}{\lower.5ex\hbox{\ltsima}}
\newcommand{\gsim}{\lower.5ex\hbox{\gtsima}}

\def\lesssim{\mathrel{\hbox{\rlap{\hbox{\lower4pt\hbox{$\sim$}}}\hbox{$<$}}}}
\def\gtrsim{\mathrel{\hbox{\rlap{\hbox{\lower4pt\hbox{$\sim$}}}\hbox{$>$}}}}

\newcommand{\beq}{\begin{equation}}
\newcommand{\eeq}{\end{equation}}
\def\beqa{\begin{eqnarray}}
\def\eeqa{\end{eqnarray}}
\def\LCDM{\ensuremath{\Lambda}CDM}

\def\head{ \vbox to 0pt{\vss \hbox to 0pt{\hskip 440pt\rm
      LA-UR-10-07069\hss} \vskip 25pt}}

\def \kms {\ifmmode  \,\rm km\,s^{-1} \else $\,\rm km\,s^{-1}  $ \fi }
\def \kpc {\ifmmode  {\rm kpc}  \else ${\rm  kpc}$ \fi  }  
\def \hkpc {\ifmmode  {h^{-1}\rm kpc}  \else ${h^{-1}\rm kpc}$ \fi  }  
\def \hMpc {\ifmmode  {h^{-1}\rm Mpc}  \else ${h^{-1}\rm Mpc}$ \fi  }  
\def \Mpch {\ifmmode  {h^{-1}\rm Mpc}  \else ${h^{-1}\rm Mpc}$ \fi  }  
\def \Msun {\ifmmode {\rm M}_{\odot} \else ${\rm M}_{\odot}$ \fi} 
\def \hMsun {\ifmmode h^{-1}\,\rm M_{\odot} \else $h^{-1}\,\rm M_{\odot}$ \fi}

\def \LCDM {\ifmmode \Lambda{\rm CDM} \else $\Lambda{\rm CDM}$ \fi}
\def \sig8 {\ifmmode \sigma_8 \else $\sigma_8$ \fi} 
\def \OmegaM {\ifmmode \Omega_{\rm m} \else $\Omega_{\rm m}$ \fi} 
\def \Omegab {\ifmmode \Omega_{\rm b} \else $\Omega_{\rm b}$ \fi} 
\def \OmegaL {\ifmmode \Omega_{\rm \Lambda} \else $\Omega_{\rm \Lambda}$\fi} 
\def \Deltavir {\ifmmode \Delta_{\rm vir} \else $\Delta_{\rm vir}$ \fi}
\def \rhocrit {\ifmmode \rho_{\rm crit} \else $\rho_{\rm crit}$ \fi}
\def \rhou {\ifmmode \rho_{\rm u} \else $\rho_{\rm u}$ \fi}
\def \zc {\ifmmode z_{\rm c} \else $z_{\rm c}$ \fi}

\def\aj{AJ}%
\def\araa{ARA\&A}%
\def\apj{ApJ}%
\def\apjl{ApJ}%
\def\apjs{ApJS}%
\def\ao{Appl.~Opt.}%
\def\aap{A\&A}%
\def\mnras{MNRAS}%
\def\pasp{PASP}%
\def\nat{Nature}%
\def\head{ .ps \vbox to 0pt{\vss \hbox to 0pt{\hskip 440pt\rm
      LA-UR-10-07069\hss} \vskip 25pt}} 

\def \spose#1{\hbox  to 0pt{#1\hss}}  
\def \lta{\mathrel{\spose{\lower 3pt\hbox{$\sim$}}\raise 2.0pt\hbox{$<$}}}
\def \gta{\mathrel{\spose{\lower 3pt\hbox{$\sim$}}\raise 2.0pt\hbox{$>$}}}

\title[The abundance bimodality of disk stars]{On the origin of the chemical bimodality of disk stars: A tale of merger and migration}

\author[T. Buck] {Tobias Buck$^{1}$\thanks{E-mail: tbuck@aip.de} \\
$^1$Leibniz-Institut f\"ur Astrophysik Potsdam (AIP), An der Sternwarte 16, D-14482 Potsdam, Germany
}

\begin{document}

\date{Accepted XXXX . Received XXXX; in original form XXXX}

\pagerange{\pageref{firstpage}--\pageref{lastpage}} \pubyear{2019}

\maketitle

\label{firstpage}

\begin{abstract}
The Milky Way's stellar disk exhibits a bimodality in the [Fe/H] vs. [$\alpha$/Fe] plane, showing a distinct high-$\alpha$ and low-$\alpha$ sequence whose origin is still under debate. We examine the [Fe/H]-[$\alpha$/Fe] abundance plane in cosmological hydrodynamical simulations of Milky Way like galaxies from the NIHAO-UHD project and show that the bimodal $\alpha$-sequence is a generic consequence of a gas-rich merger at some time in the Galaxy's evolution. The high-$\alpha$ sequence evolves first in the early galaxies, extending to high metallicities, while it is the low-$\alpha$ sequence that is formed after the gas-rich merger. The merger brings in fresh metal-poor gas diluting the interstellar medium's metallicity while keeping the [$\alpha$/Fe] abundance almost unchanged. The kinematic, structural and spatial properties of the bimodal $\alpha$-sequence in our simulations reproduces that of observations. In all simulations, the high-$\alpha$ disk is old, radially concentrated towards the galaxy's center and shows large scale heights. In contrast, the low-$\alpha$ disk is younger, more radially extended and concentrated to the disk mid-plane. Our results show that the abundance plane is well described by these two populations that have been distributed radially across the disk by migration: at present-day in the solar neighbourhood, low-$\alpha$ stars originate from both the inner and outer disk while most of the high-$\alpha$ stars have migrated from the inner disk. We show that age dating the stars in the [Fe/H]-[$\alpha$/Fe] plane can constrain the time of the low-$\alpha$ sequence forming merger and conclude that $\alpha$-bimodality is likely a not uncommon feature of disk galaxies.
\end{abstract}

\noindent
\begin{keywords}

Galaxy: structure --- galaxies:
  evolution --- galaxies: kinematics and dynamics --- galaxies:
  formation --- Galaxy: disk --- methods: numerical
 \end{keywords}


\section{Introduction} \label{sec:introduction}

Since \citet{Gilmore1983} first separated the stellar disk of the Milky Way (MW) into two distinct components of different scale height and age, it has been recognized that such a distinct structure is ubiquitous in the local universe \citep{Yoachim2006}. Subsequently, it was realized that this dichotomy in vertical disk structure might encode essential information about the formation of the major baryonic component of our MW in particular and disk galaxies in general. By now several different mechanisms for creating structurally distinct thick disks exist in the literature ranging from vertical disk heating by satellite encounters \citep{Quinn1993,Villalobos2008}, the accretion of satellite stars \citep{Abadi2003}, gas-rich mergers \citep{Brook2004}, major mergers \citep{Belokurov2018,Helmi2018} or star formation in turbulent gaseous disks at high redshift \citep{Noguchi1998,Bournaud2009}. Also a purely secular formation mechanism was proposed by invoking the radial migration of kinematically hot stars from the inner to the outer disk \citep{Schönrich2009,Loebman2011,Roskar2012} which is expected to be important to reproduce the flat age-metallicity relation \citep{Casagrande2011} and the metallicity-rotation velocity relation \citep{Allende-Prieto2016,Kordopatis2017,Schönrich2017}. However, it is still a matter of debate whether radial migration actually contributes to the formation of spatially thick disks \citep{Minchev2012,Kubryk2013,Vera-Ciro2014,Grand2016,Kawata2017} or if these simply reflect the inside-out growth of stellar disks \citep{Minchev2015}.

The idea of two distinct disk components is further supported by a separation of solar neighbourhood disk stars in the [$\alpha$/Fe] vs. [Fe/H] plane 
\citep{Gratton1996,Fuhrmann1998}. 
Here $\alpha$ refers to the $\alpha$-elements which are mainly produced in core-collapse supernovae (CC-SN) while iron is mainly produced in supernova type Ia (SNIa) explosions.
The progenitor stars of the former have a short lifetime ($\lesssim40$ Myr) compared to the progenitors of SNIa. Thus, the timescale of chemical enrichment in the ISM is different for these two element groups. Depending on the relative contribution of recycled gas from CC-SN and SNIa to the ISM the next generation of stars formed can inhabit different space in the [Fe/H] vs. [$\alpha$/Fe] space \citep[see e.g.][for more details]{Nomoto2013}. Large contributions of CC-SN as e.g. after a burst of star formation in pristine gas will mainly lead to the production of $\alpha$ elements and thus to $\alpha$ enhanced stars in the next generation. On the other hand, a larger contribution of SNIa would mainly add iron to the ISM and could bring down the [$\alpha$/Fe] ratio of already $\alpha$ enhanced gas to form low-[$\alpha$/Fe] stars. In this way star formation, gas accretion and chemical enrichment may form complex patterns in chemical abundance space of the disk stars encoding evolutionary information of the galaxy.     

Galactic spectroscopic surveys such as e.g. RAVE \citep{Steinmetz2006}, APOGEE \citep{apogee}, GALAH \citep{Galah,Buder2018} or LAMOST \citep{lamost} have now confirmed and further extended the dichotomy of the stellar disk in kinematics, age and chemistry. By now it is well established that the local stellar disk separates into a high-$\alpha$ and a low-$\alpha$ disk over several orders of magnitude in stellar iron abundance where the high-[$\alpha$/Fe] component is thicker and more radially compact than the low-[$\alpha$/Fe] component \citep{Reddy2003,Bensby2011,Bovy2012,Anders2014,Hayden2015}. These findings further complicate the interpretation of the thick-thin disk dichotomy as there is no simple mapping between the thick disk as defined in real space into the thick, $\alpha$ enhanced disk. Slicing instead the stellar disk into stellar sub-populations of narrow ranges in iron abundance (mono abundance populations, MAPs), \citet{Bovy2016} found the surface densities of low-$\alpha$ MAPs to consist of "donut"-like structures while the ones of the high-$\alpha$ components are centrally concentrated. While the scale radii and scale heights of the high-$\alpha$ MAPs are roughly constant, the low-$\alpha$ "donuts" become radially larger and vertically thicker with decreasing metallicity. Similar findings also hold if the disk is instead sliced into mono-age populations \citep{Minchev2017,Mackereth2017}. Together, this points towards a different formation scenario for the two chemically defined disks \citep{Haywood2015}.

\citet{Chiappini1997} were the first to analytically tie together the properties of a geometrically and chemically distinct disk by two phases of star formation. In this scenario the thick disk is formed from an early rapid star formation phase while the thin disk forms out of the gradual infall of primordial gas \citep{Spitoni2009,Spitoni2019}. This model can further explain the dichotomy in the local metallicity distribution, under the assumption that metal-rich, high-[$\alpha$/Fe] stars migrated from the inner to the outer disc \citep[but see also][for an alternative explanation]{Grisoni2017}. Still, these models struggle to explain the extremely metal-rich stars of broad ranges in age which obey thin disk kinematics at the local volume \citep{Trevisan2011,Anders2017}.

With the advent of state-of-the art simulations with enough resolution to study the vertical structure of galactic disks \citep[e.g.][]{Obreja2016,Ma2017,Grand2017,Buck2019d} the formation scenario of chemically and geometrically distinct disks can be studied self-consistently in fully cosmological context. Such simulations predict that thin stellar disks form in an inside-out, upside-down fashion \citep[e.g.][]{Stinson2013,Bird2013,Minchev2014,Grand2016a,Dominguez-Tenreiro2017} while centrally concentrated, thick components are formed by violent gas rich mergers \citep{Brook2004,Brook2007}. However, the exact formation mechanism is still under debate and reproducing the chemical dichotomy poses still a challenge for modern simulations. The question arises if a chemical bimodality is at all an ubiquitous feature of disk galaxies. For example, simulations of \citet{Brook2012} do find chemically distinct disks \citep[see also][]{Snaith2016} whereas chemical abundance space is homogeneous in \citet{Minchev2013} and in the AURIGA galaxies both possibilities are realized \citep{Grand2018}. To complicate the situation further, different formation scenarios have been suggested in the literature ranging from a two infall scenario \citep[e.g.][]{Chiappini1997,Grisoni2017,Spitoni2019}, a central star burst \citep[e.g.][]{Grand2018} or merger scenario \citep[e.g.][]{Calura2009}. And recently \citet{Clarke2019} showed that a chemical bimodality is formed by vigorous star formation in clumps at high redshift. Thus, the formation of chemically distinct disks in cosmological simulations is still an unsettled question with strong implications for the interpretation of the observed chemical dichotomy of MW's stellar disk.

Here we use the NIHAO-UHD suite of cosmological zoom-in simulations of MW mass galaxies \citep{Buck2019d} to study the structure and formation of the stellar disk in chemical abundance space. In \S2 we introduce the simulations and summarize simulation parameters. In \S3 we study the evolution of the [$\alpha$/Fe] vs. [Fe/H] plane as a function of radius and vertical height from the disk mid-plane. In three out of four cases we find clearly defined sequence in the abundance plane. In \S4 we investigate the origin of the chemical dichotomy by tracing the time evolution of the gas mass and gas iron and oxygen abundance as a function of time. We connect this to the age and birth radius structure in the [$\alpha$/Fe] vs. [Fe/H] to establish the merger and migration origin of the chemically distinct disks in our simulations. We conclude this work in \S5 with a short summary and discussion of our findings in comparison to observational results for the MW.

\begin{figure*}
\begin{center}
\includegraphics[width=\textwidth]{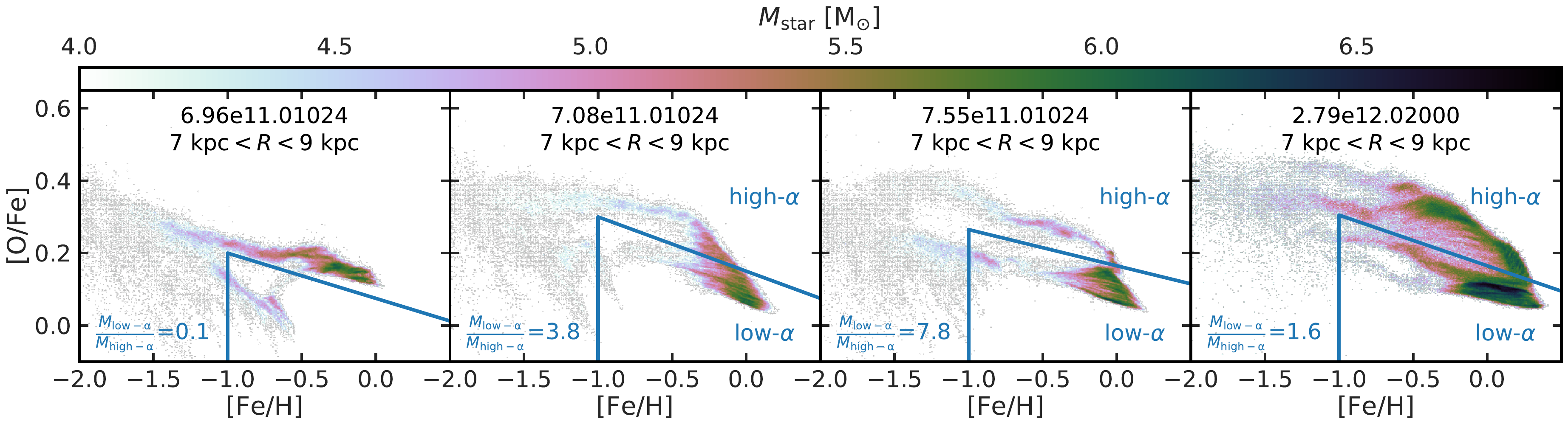}
\end{center}
\vspace{-.35cm}
\caption{[$\alpha$/Fe] vs. [Fe/H] plane at the solar radius for our four high-resolution simulations. All four simulations clearly show two separate sequences at different oxygen abundances over a large range in iron abundance but only in three of the simulations the low-$\alpha$ sequence extends all the way to [Fe/H]$>0.0$. With the blue box we show our separation of the low-$\alpha$ from the high-$\alpha$ sequence and quote the stellar mass ratio of the two sequences in the lower left corner of each panel.}
\label{fig:impression}
\end{figure*}

\newpage
\section{Cosmological Zoom-in Simulations} \label{sec:simulation}

For this work we use four simulations from the NIHAO-UHD 
suite \citep{Buck2019d} part of the Numerical Investigation of a Hundred Astronomical Objects (NIHAO) simulation suite \citep{Wang2015}.
\citet{Buck2019d} compared in detail the properties of these galaxies to observations of the MW and local disk galaxies from the SPARC data \citep{Lelli2016} showing that simulated galaxy properties agree well with observations. Of particular interest for this study are the stellar disk size, vertical scale heights as well as stellar disk masses which are all shown to be in good agreement with observations of the MW. Parts of this simulation suite have previously been used to study the build-up of MW's peanut-shaped bulge \citep{Buck2018,Buck2019b} or the dwarf galaxy inventory of MW mass galaxies \citep{Buck2019c}.

The simulations assume cosmological parameters from the \cite{Planck}, namely: \OmegaM= 0.3175, \OmegaL= 0.6825,
\Omegab= 0.049, H${_0}$ = 67.1\kms\Mpc$^{-1}$, \sig8 = 0.8344. Initial conditions are created the same way as for the original NIHAO runs \citep[see][]{Wang2015} using a modified version of the \texttt{GRAFIC2} package \citep{Bertschinger2001,Penzo2014}. The mass resolution of these simulation ranges between $m_{\rm dark}\sim1.5 - 5.1\times10^5 \Msun$ for dark matter particles and $m_{\rm gas}\sim2.8 - 9.4\times10^4 \Msun$ for the gas particles. Stellar particles are born with an initial mass of $1/3\times m_{\rm{gas}}$ and are subject to massloss according to the stellar evolution models. The corresponding force softenings are $\epsilon_{\rm dark}=414 - 620$ pc for the dark matter and $\epsilon_{\rm gas}=177 - 265$ pc for the gas and star particles. However, the adaptive smoothing length scheme implies that $h_{\rm  smooth}$ can be as small as $\sim30$ pc in the disk mid-plane.
The simulation setup, star formation and feedback implementations are described in detail in the introductory paper \citep{Buck2019d} but for completeness we summarise them below. 

Simulations are performed with the modern smoothed particle hydrodynamics (SPH) solver {\texttt{GASOLINE2}} \citep{Wadsley2017} including substantial updates to the hydrodynamics as described in \citet{Keller2014}. Cooling via hydrogen, helium, and various  metal-lines is implemented as described in \citet{Shen2010} using \texttt{cloudy} \citep[version 07.02;][]{Ferland1998} tables. These calculations include photo-heating from the \citet{Haardt2005} UV background\footnote{For details on the impact of the UV background on galaxy formation see the recent study by \citet{Obreja2019}}. Star formation proceeds in cold (T $< 15,000$K), dense ($n_{\rm  th}  >  10.3$cm$^{-3}$) gas and is implemented as described in \citet{Stinson2006}. In \citet{Buck2019} it was shown that in these kind of simulations only a high value of $n_{\rm  th}>10$cm$^{-3}$ \citep[see also][for an extended parameter study]{Dutton2019} is able to reproduce the clustering of young star clusters observed in the Legacy Extragalactic UV Survey (LEGUS) \citep{Calzetti2015,Grasha2017}.

Two modes of stellar feedback are implemented following \citet{Stinson2013}. The  first  mode models the energy input from young massive stars, e.g. stellar  winds and photo ionization. This mode happens before any supernovae explode and consists of the total stellar luminosity  ($2 \times 10^{50}$  erg of  thermal energy  per $M_{\odot}$) of the entire stellar population. The efficiency parameter for coupling the energy input is set to $\epsilon_{\rm ESF}=13\%$ \citep{Wang2015}. The second mode models supernova explosions and is implemented using the blastwave formalism as described in \citet{Stinson2006}. This mode applies a delayed cooling formalism for particles inside the blast region following \citet{McKee1977} in order to account for the adiabatic expansion of the supernova.
Finally, most important for this study, we adopted a metal diffusion algorithm between particles as described in \citet{Wadsley2008}. Throughout this paper we show results for all stars within a cylinder of height $\vert z\vert<5$ kpc from the stellar disk which we assume to be in the x-y-plane. Following the nomenclature used in observations and acknowledging the fact that the oxygen abundance in the simulations is more closely tracing the $\alpha$ element abundance, from now on we refer to it as $\alpha$ and with that to the two sequences as the high- and low-$\alpha$ sequence, respectively. 

\begin{figure*}
\begin{center}
\includegraphics[width=\textwidth]{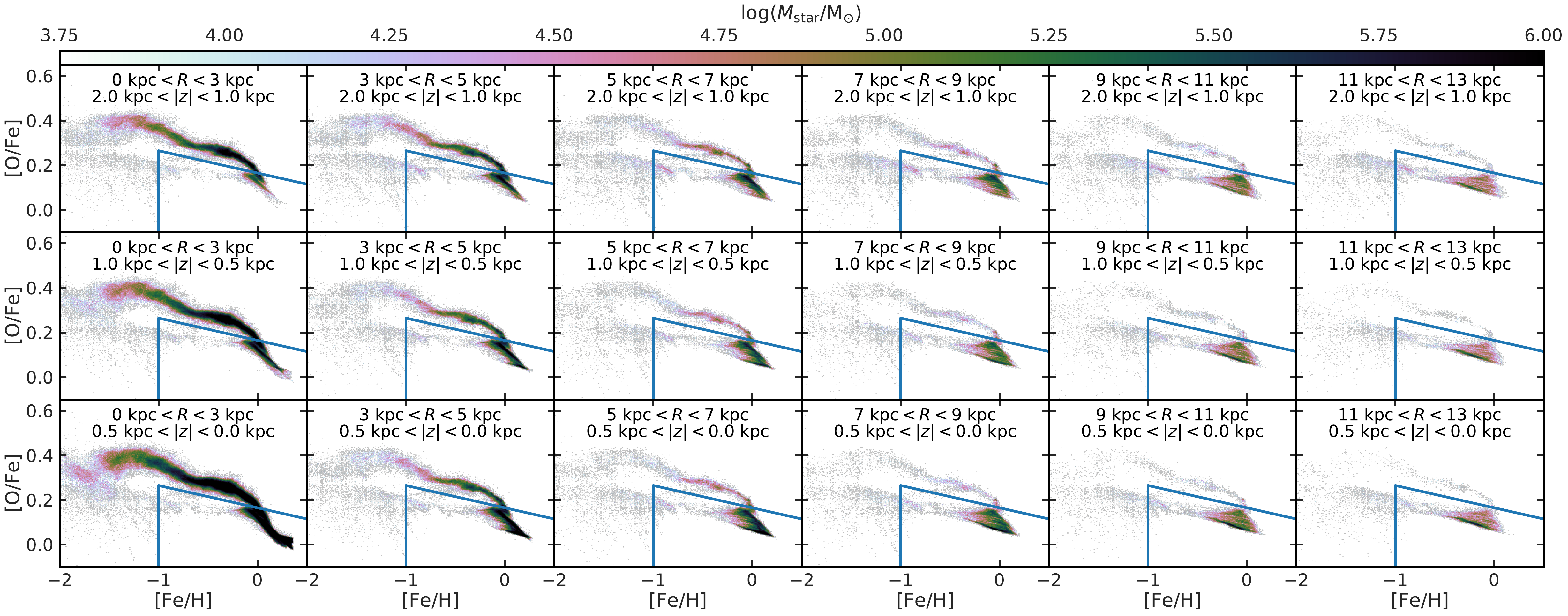}
\end{center}
\vspace{-.35cm}
\caption{[$\alpha$/Fe] vs. [Fe/H] plane for the galaxy g7.55e11 in six different galacto-centric annuli of increasing radius from left to right as indicated in the panels. From top to bottom we show stars in three bins of decreasing height from the galactic mid-plane. In each panel the blue box shows our separation into low- and high-$\alpha$ sequence as introduced in figure \ref{fig:impression}. We qualitatively recover the observational fact that the strength of the high-$\alpha$ sequence decreases with increasing galacto-centric radius and increases with larger heights from the galactic mid-plane \citep[e.g.][]{Hayden2015}. The low-$\alpha$ sequence on the other hand becomes more prominent at larger radii and diminishes at large heights above the mid-plane.}
\label{fig:hayden}
\end{figure*}

\section{Chemical bimodality in the [$\alpha$/Fe] vs. [Fe/H] plane}
\label{sec:}

As we have discussed in the introduction, one of the most striking structural features of the MW's stellar disk is its bimodality in the [$\alpha$/Fe] vs [Fe/H] plane. In this section we qualitatively compare the structure of the simulated stellar disks to results from the MW. In figure \ref{fig:impression} we show from left to right  the [$\alpha$/Fe] vs [Fe/H] plane for disk stars ($\vert z\vert<5$ kpc) at the solar radii ($7<R<9$ kpc) in the four simulations as indicated in each panel. The color coding shows how much stellar mass falls in each pixel indicated by the colorbar on the top of this figure. In each panel we indicate the separation into the high- and low-$\alpha$ sequence by the blue lines and quote the stellar mass ratio of the two sequences in the lower left corner. This figure shows that the NIHAO-UHD simulations produce a bimodality in chemical abundance space. Sometimes we can even find less pronounced third and fourth sequences (e.g. for g7.55e11 or g2.79e12). Galaxies g2.79e12, g7.55e11 and g7.08e11 show two parallel sequences at different oxygen abundance over a large range in iron abundance for almost all galactic annuli presented here. Galaxy g6.96e11 on the other hand shows two more diagonal sequences.

Another interesting feature seen in this plot are the nearly diagonal ridges at [Fe/H]$\lesssim-1$ and the broadening of the sequences at very low metallicities. such features are not necessarily observed in the MW. In the simulations, these structures originate from mergers bringing in stars formed in satellite galaxies. That indeed satellites of the MW show different structures in the [$\alpha$/Fe] vs [Fe/H] plane has recently been shown in e.g. \citet{Nidever2019}. Their results combined with our findings here might offer another possibility to distinguish accreted stars from stars formed inside the MW, next to solely using stellar kinematics. However, such a study is outside the scope of this work and we leave it for a future work.

Additionally to the high-resolution simulations presented here, we have further investigated the lower resolution NIHAO galaxies \citep{Wang2015} with halo masses between $2\times 10^{11}-3\times10^{12}\Msun$. In 16 out of 35 cases we find a bimodality in [Fe/H] vs. [$\alpha$/Fe] space (see fig. \ref{fig:nihao} in the appendix) with the incidence of the bimodality strongly increasing with halo mass. We find some diversity in the patterns in [Fe/H] vs. [$\alpha$/Fe] space. Some of the patterns at halo masses greater than $\sim5\times10^{11}\Msun$ resemble well the patterns observed in the MW whereas others have differently shaped morphologies or no bimodality at all. Thus, a [$\alpha$/Fe] bimodality is a generic feature in our simulations unlike results from the EAGLE simulations \citep[e.g.][]{Mackereth2019} or the AURIGA suite \citep{Grand2018}. These differences might be due to the sub-grid treatment of the ISM which results in a relatively smooth ISM in the latter case \citep[see fig. e.g.][]{Marinacci2019} compared to the simulations used in this work which lead to the formation of clumpy disks at higher redshifts \citep[$z\gtrsim1$][]{Buck2017} in agreement with observations \citep[e.g.][]{Guo2015,Shibuya2016}.

\subsection{Spatial variations in the [Fe/H] vs. [$\alpha$/Fe] plane}
At closer inspection, we find that there is a strong evolution of the sequence strength with galacto-centric radius and vertical height from the disk mid-plane for all simulations. This behaviour is shown in fig. \ref{fig:hayden} for model galaxy g7.55e11. From top to bottom we show stars inhabiting three different bins in vertical height above the disk mid-plane and from left to right we show six different radial ranges as indicated iwith the labels in each panel. The stellar mass contained in the high-$\alpha$ sequence decreases with increasing galacto-centric radius sometimes completely vanishing beyond $R\sim10$ kpc. At the same time the low-$\alpha$ sequences is more pronounced in the galactic outskirts and first clearly appears at $R\sim5$ kpc. Furthermore, at fixed radius the high-$\alpha$ sequence stars exhibit on average larger distances from the disk mid-plane while the low-$\alpha$ sequence is more concentrated towards the disk mid-plane. Such an evolution of the two sequences matches the observed behaviour in the MW \citep[see e.g.][]{Hayden2015}.

\subsection{Stellar ages and birth radii in the [Fe/H] vs. [$\alpha$/Fe] plane}

\begin{figure*}
\begin{center}
\includegraphics[width=\textwidth]{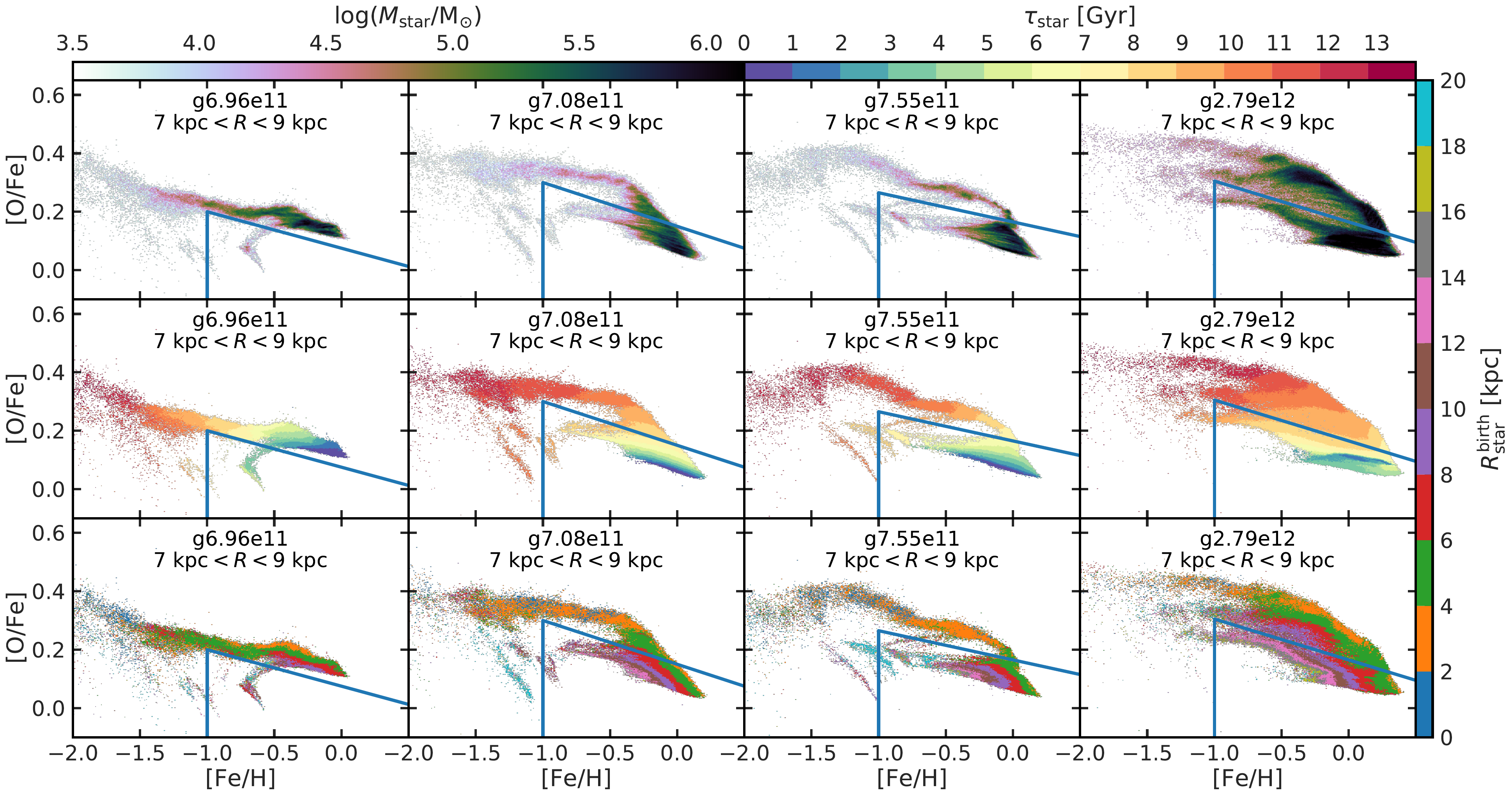}
\end{center}
\vspace{-.35cm}
\caption{[$\alpha$/Fe] vs. [Fe/H] plane at the solar radius ($7<R<9$ kpc) for all star particles born inside the galaxy (excluding accreted star particles). From left to right we show all four simulations and from top to bottom we color code the plane by stellar mass, stellar age and birth radius as indicated in the different colorbars. This figure shows an exceptionally old high-$\alpha$ sequence whose stars were mostly born in the center of the galaxies and must have migrated to the solar circle afterwards. In contrast to this, the low $\alpha$ sequence is formed more recently within the last $7$ Gyr and consist of stars with a large range of birth radii, both inside and outside the solar circle. Figures for all radii are shown in the appendix.}
\label{fig:insitu}
\end{figure*}

In order to investigate the origin of the double sequence, in figure \ref{fig:insitu} we focus on an annulus between $7<R<9$ kpc matching the solar vicinity. We further exclude all stars accreted in satellites which is less than a few percent for the galaxies under consideration here \citep{Buck2019d}.
From left to right we show all four simulations and from top to bottom we color code the [$\alpha$/Fe] vs [Fe/H] plane by stellar mass (same as in figure \ref{fig:impression} in order to guide the eye on how much stellar mass is found in each pixel), stellar age and by birth radius. A similar figure for all radii bins and each of the four galaxies is shown in the appendix (\ref{fig:app1} and \ref{fig:app2}). 

\subsubsection{Stellar ages}
Looking at the middle row of average stellar ages we find that there is a clear but continuous horizontal trend of decreasing stellar age with decreasing $\alpha$ abundance for both the high-$\alpha$ and low-$\alpha$ sequence. This is remarkably similar to the observed trends in the MW where the low-$\alpha$ sequence exhibits younger stars compared to the high-$\alpha$ sequence \citep[e.g. fig. 2 of][and fig. 7 of \citealt{Feuillet2019}]{Ness2019}. On the other hand this leads to the conclusion that the enrichment history early on in the formation of these galaxies is a rapid process such that stellar ages along the high-$\alpha$ sequence are very similar and exclusively old (see also fig. \ref{fig:age}). After the knee when supernovae-Ia bring down the $\alpha$ abundance, the chemical evolution slows down and the former appears to be a good predictor of stellar age. The low-$\alpha$ sequence is comparatively younger than the high-$\alpha$ one and we find an age gradient in both directions in the [$\alpha$/Fe] vs. [Fe/H] plane in excellent agreement with observations from the MW \citep[see fig. 2][]{Ness2019}. Thus, the NIHAO-UHD simulations predict a large age gradient at fixed [Fe/H] across [$\alpha$/Fe] and at large metallicities (mostly for the low-$\alpha$ stars) we predict a slight age gradient at fixed [$\alpha$/Fe] across [Fe/H]. Similar age trends are also found by \citet{Grand2018} although with a different slope for the coeval populations.

The sharp age contrast in the middle panels of fig. \ref{fig:insitu} suggest, that the low-$\alpha$ sequence starts at a specific point in time and then co-evolves with the high-$\alpha$ sequence in these simulations. This points towards a relatively quick change in ISM metallicity at that point in time when the low-$\alpha$ sequence first appeared. At that time the ISM iron abundance must have been diluted to lower values from which the chemical evolution then continues in order to form the low-$\alpha$ sequence. 
Thus, the scenario presented here leads to a purely old high-$\alpha$ sequence and a younger low-$\alpha$ sequence as it is observed in the MW \citep[e.g.][]{Bensby2014,Buder2019}. In contrast to our findings, the recent study by \citet{Clarke2019} invokes a scenario in which star forming clumps at high redshift are the origin of the high-$\alpha$ sequence. These clumps start out from the low-$\alpha$ sequence and quickly self-enrich in order to form the high-$\alpha$ sequence. Thus, comparing ages of the two sequences in this scenario, one would expect to find relatively younger stars in the high-$\alpha$ sequence compared to the low-$\alpha$ sequence at fixed metallicity which is actually not observed.

\begin{figure}
\begin{center}
\includegraphics[width=\columnwidth]{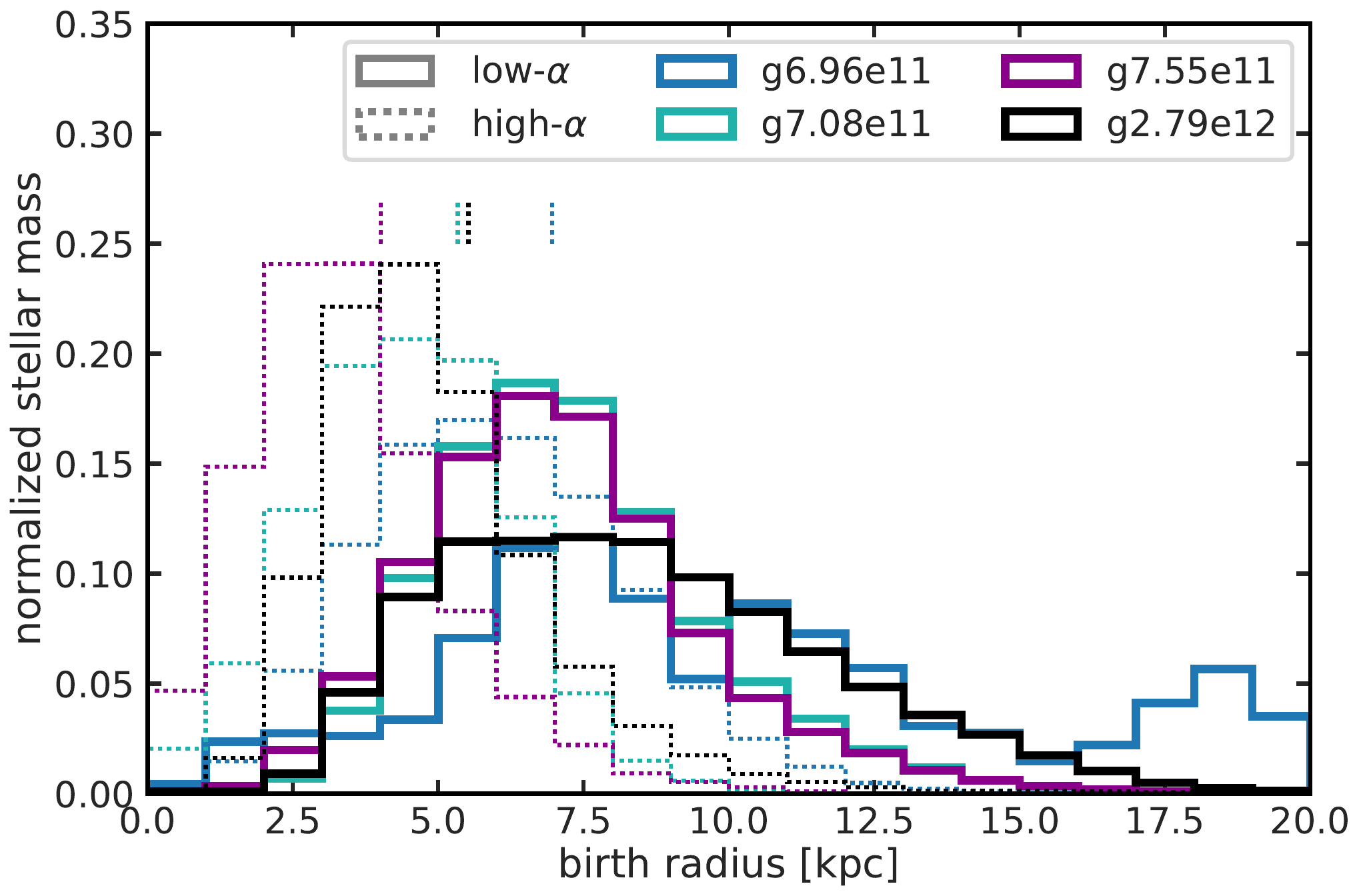}
\end{center}
\vspace{-.35cm}
\caption{Histograms of stellar birth radii in the low-$\alpha$ sequence (solid lines) and the high-$\alpha$ sequence (dotted lines) for the four simulations. Vertical dotted lines indicate the 68th percentile of the birth radii distribution of the high-$\alpha$ stars. Stars in the high-$\alpha$ sequences are mostly born in the inner disk ($R_{\rm birth}\lesssim6$ kpc).}
\label{fig:birth_r}
\end{figure}

\begin{figure}
\begin{center}
\includegraphics[width=\columnwidth]{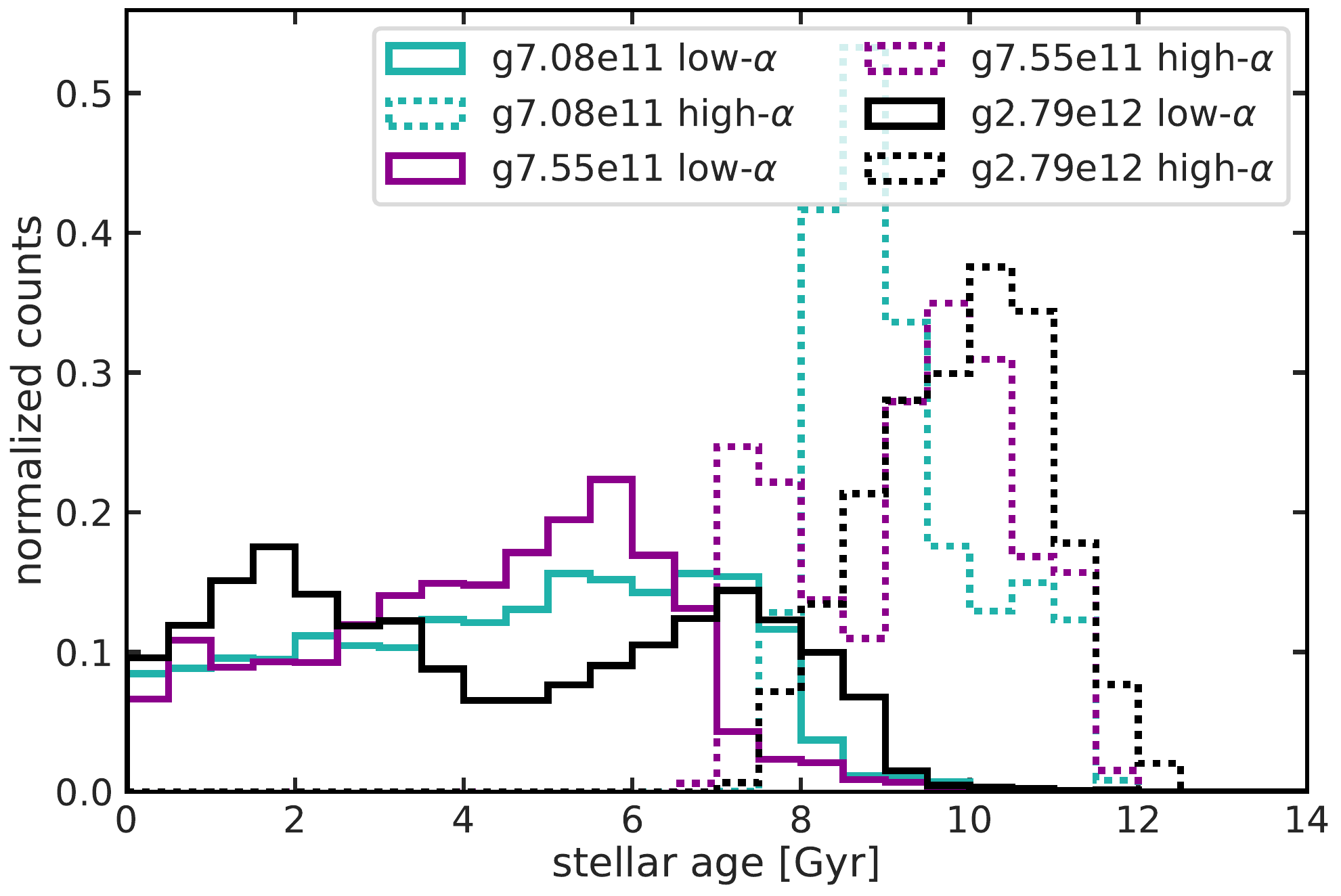}
\end{center}
\vspace{-.35cm}
\caption{Histograms of stellar ages in the low-$\alpha$ sequence (solid lines) and the high-$\alpha$ sequence (dotted lines) for the three simulations with a strong abundance bimodality. Stars in the high-$\alpha$ sequences are exclusively old while low-$\alpha$ stars are younger but show a tail towards larger ages owing to the overlap of high- and low-$\alpha$ sequences at the highest metallicities.}
\label{fig:age}
\end{figure}

In order to make this point clearer, fig. \ref{fig:age} shows the age distributions of low- and high-$\alpha$ disk stars for the three simulations which show a clear bimodality. The solid histograms show the low-$\alpha$ sequence stars, the dotted histograms the high-$\alpha$ stars. This figure confirms our previous findings that the two sequences are also distinct in stellar ages. The high-$\alpha$ disk stars are exclusively old while low-$\alpha$ stars are younger with a tail towards ages of $\sim7-10$ Gyr. This tail of older stars is the result of the overlap of the two sequences at the highest metallicities. Comparing our results to observations for the MW \citep[e.g. fig. 9 of][]{Silva2018} we find good agreement. However, we do not see a tail of young high-$\alpha$ stars in our simulations which might point towards miss-labeled ages in the observations \citep[e.g.][]{Martig2015,Yong2016,Jofre2016}. Another origin of these stars was proposed by \citet{Chiappini2015} and attributed to migrators from the Galactic bar. However, our simulations include one strongly barred galaxy (g2.79e11) which also does not show any young high-$\alpha$ stars. On the other hand, if the observed stars turn out to be truly young they might help to better constrain the employed simulation prescriptions.

\subsubsection{Birth radii}
Finally, the lower panels in fig \ref{fig:insitu} color code the birth radius of the stars. We accompany this by a histogram of the birth radii (Fig. \ref{fig:birth_r}) of stars in the high- and low-$\alpha$ sequence as defined by the blue box. This shows, that the high-$\alpha$ stars mostly originate from the inner galaxy ($R\lesssim6$ kpc) while the lower sequence separates into a range of birth radii where larger birth radii are offset to lower metallicity. Only galaxy g6.96e11 without a clear low-$\alpha$ sequence shows a large fraction of high-$\alpha$ stars born in the solar neighbourhood. In general, the lower sequence traces very well the knee where the split in birth radii follows the shape of the knee. This points towards a self-similar chemical evolution at each radius whereby the exact position in the [Fe/H] vs. [$\alpha$/Fe] plane is simply set by the gradients of iron and $\alpha$ abundance where in general the simulations show lower iron abundance and higher $\alpha$ abundance in the outskirts.

Bringing the findings of this figure together, the horizontal age gradient and the "diagonal" radius separation in the [Fe/H] vs. [$\alpha$/Fe] is simply a reflection of the fact that to first order at each time star formation in the disk happens at different radii whose metallicity decreases with increasing radii due to the metallicity gradient of the disk itself. However, that we find all these birth radii in the solar vicinity tells us a lot about the impact of radial migration in these disks. All panels in fig. \ref{fig:insitu} are selected to only show stars in between $7<R<9$ kpc but we find that the high-$\alpha$ sequence stars must have migrated outwards in order to be found at $\sim8$ kpc today while for the low-$\alpha$ sequence we find that stars come both from the inner and the outer disk (sometimes as far as $15-20$ kpc). These results are in agreement with estimates of the radial migration strength of the MW disk stars \citep[$\sim3.6$ kpc][]{Frankel2018} and recent findings of a strong impact of radial migration in the MW by \citet{Feuillet2019}. We find that the $\alpha$-bimodality is already present at time of birth of the stars but with a much stronger radial separation. At birth high-$\alpha$ stars were more concentrated to the galactic center while low-$\alpha$ stars were born further out. Radial migration washes out this radial separation and leads to a stronger overlap of the two sequences at the solar radius at present-day. 

To conclude, Figs. \ref{fig:insitu}, \ref{fig:birth_r} and \ref{fig:age} show that during the early history of these MW analogues chemical evolution quickly pre-enriches the ISM to form the high-$\alpha$ sequence. From there the evolution then progresses at each radius almost self-similar reflecting the metal gradient in the disk. Then, at some point in time a rapid event leads to the dilution of the iron abundance in the ISM while at the same time keeping the $\alpha$ abundance almost constant. Thus, the next generation of stars will be born at lower metallicity but the same [$\alpha$/Fe] value as the previous generation and as such forming the low-$\alpha$ sequence. In this way, a chemical bimodality of the disk stars (at each radius) is established at the time of birth of the low-$\alpha$ stars. This bimodality is then modified by radial orbit migration during the successive evolution of the galaxy. As we will show in the next section, such a dilution is caused by gas rich mergers of satellites in these simulations. The almost pristine gas of the satellites will bring in a lot of hydrogen therefore lowering the [Fe/H] ratio while almost no iron or at least the same amount of iron and $\alpha$ is brought into the disk ISM by this event. Thus, the ratio of [$\alpha$/Fe] stays almost constant. Here it is only important that the merger is gas rich in order to dilute the disk gas but not a major merger in which case it might destroy the stellar disk.

\subsection{Origin of the chemical bimodality}

\begin{figure}
\begin{center}
\includegraphics[width=.925\columnwidth]{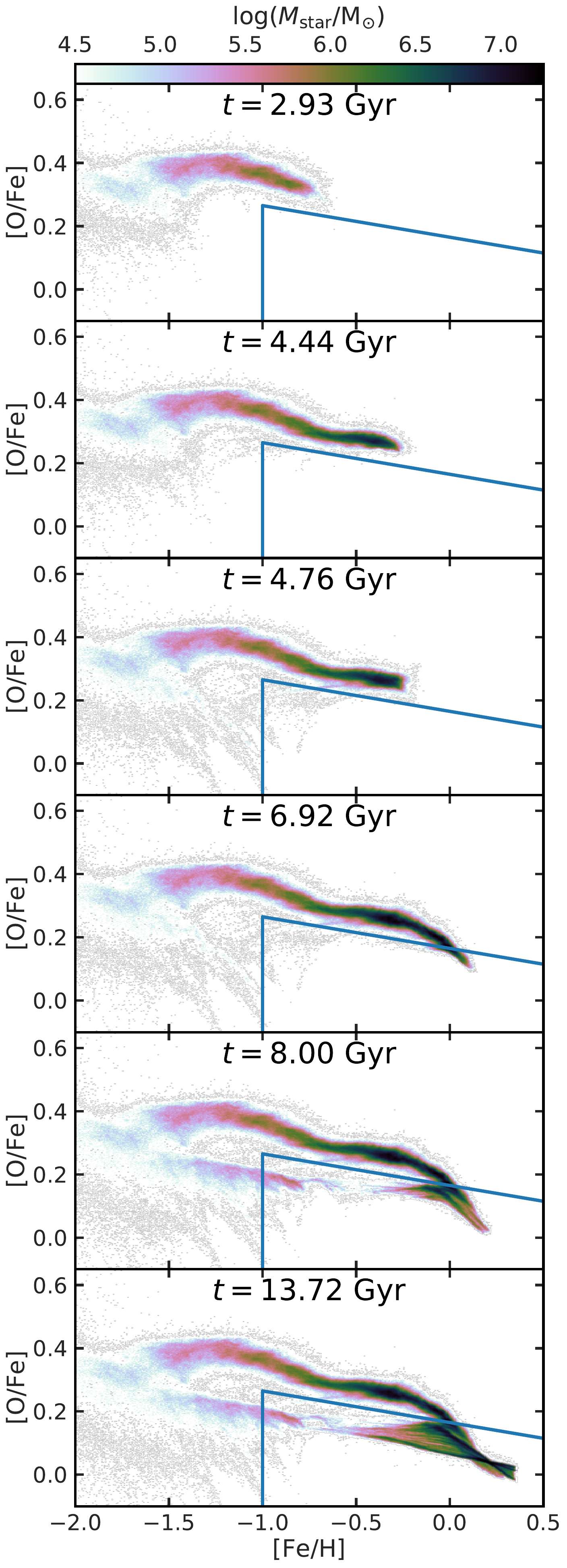}
\end{center}
\vspace{-.35cm}
\caption{Time evolution of the [$\alpha$/Fe] vs. [Fe/H] plane for g7.55e11 for all star particles inside the galaxy at different cosmic times increasing from top to bottom. Chemical evolution quickly enriches the ISM to high values of [$\alpha$/Fe] at low [Fe/H]. From there the high-$\alpha$ sequence slowly evolves to high metallicities and lower [$\alpha$/Fe]. Around $t\sim7$ Gyr a gas rich merger dilutes the ISM (see text for more explanation) and the low-$\alpha$ sequence forms.}
\label{fig:time}
\end{figure}

\begin{figure*}
\begin{center}
\includegraphics[width=\textwidth]{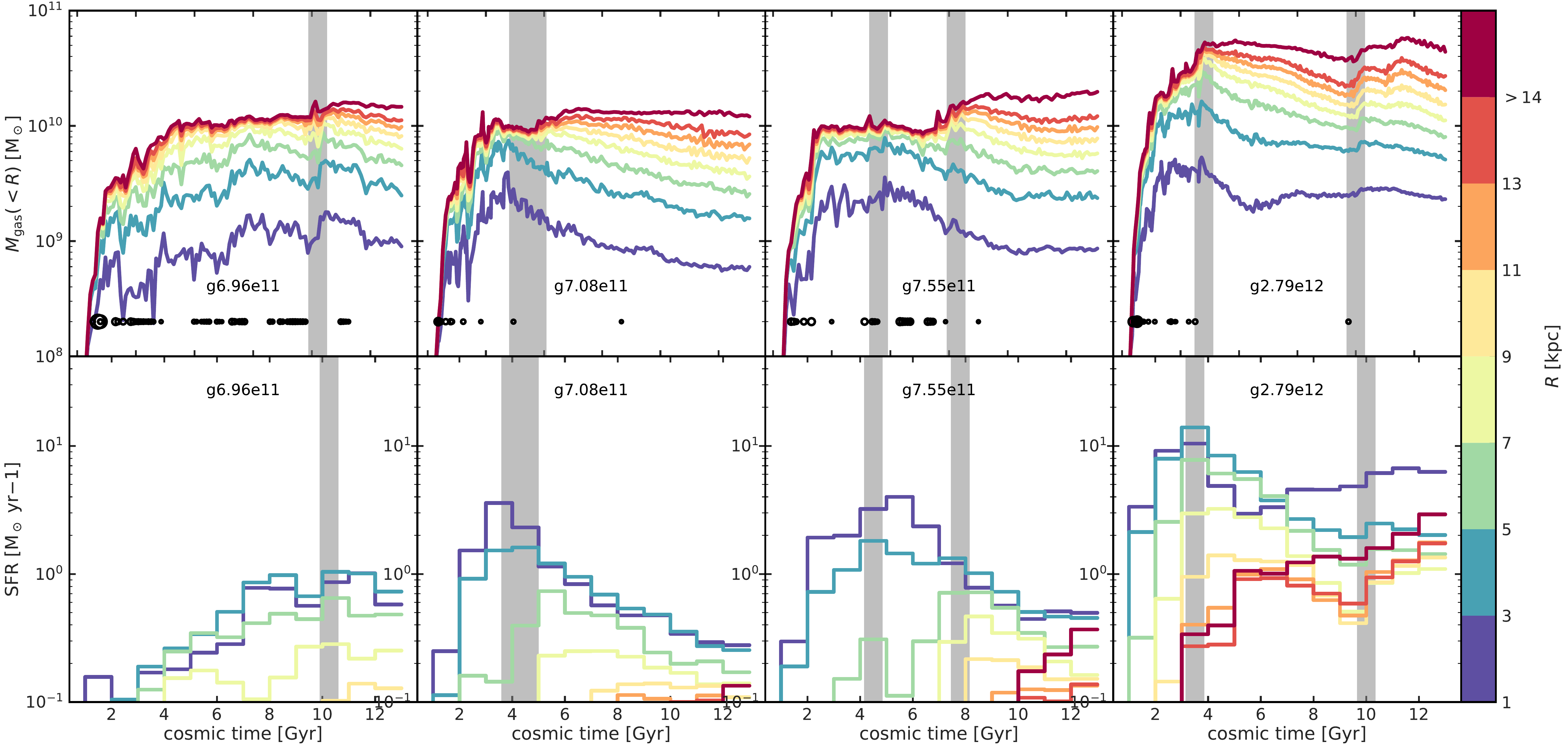}
\end{center}
\vspace{-.35cm}
\caption{\emph{Upper panel}: Disk gas mass inside different galacto-centric radii $R$ indicated by the line color and the colorbar on the right as a function of time. \emph{Lower panel}: Star formation history in different galacto-centric annuli of width $2$ kpc centred on the radii indicated by the colorbar on the right. Small black dots highlight times of gas rich merger with the gray bands highlighting the point in time when the lower $\alpha$ sequence first emerged. Dot sizes show the fractional gas mass contribution of each merger to the main galaxy's gas reservoir. From left to right we show the different simulations as indicated with labels in the panels.}
\label{fig:gas_mass}
\end{figure*}

\begin{figure*}
\begin{center}
\includegraphics[width=\textwidth]{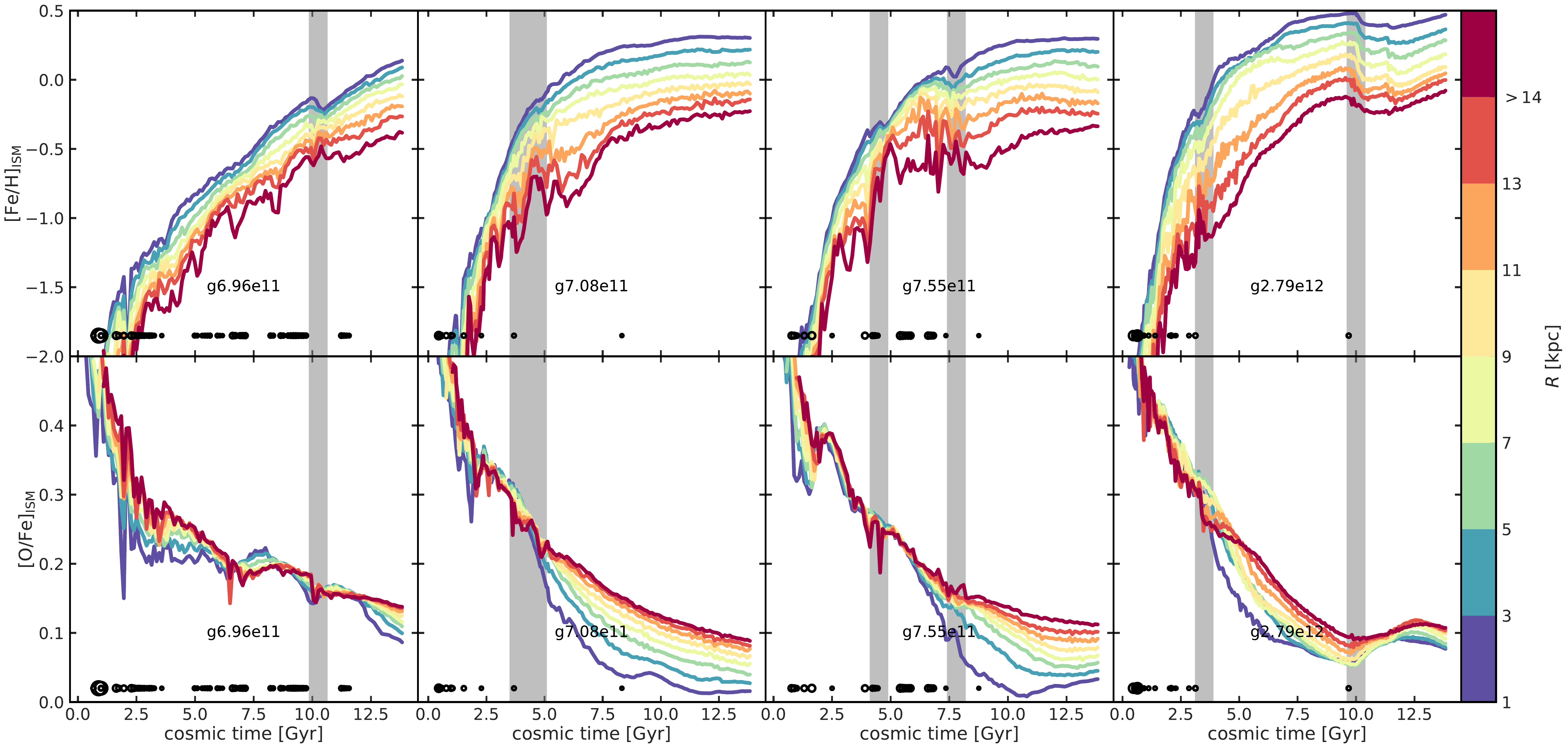}
\end{center}
\vspace{-.35cm}
\caption{Time evolution of the iron (upper panel) and oxygen (lower panel) abundance. We show different galacto-centric radii as indicated by the colorbar on the right. Small black dots again highlight times of gas rich mergers and size give the fractional gas mass contribution to the galaxy's gas reservoir. Gray bands highlighting the point in time when the lower $\alpha$ sequence first emerged. From left to right we show the different simulations as indicated with labels in the panels.}
\label{fig:ISM}
\end{figure*}

In this section we investigate in detail how the low-$\alpha$ sequence is formed by gas rich mergers in the simulations. To that extent, in figure \ref{fig:time} we show from top to bottom a time evolution of the [$\alpha$/Fe] vs. [Fe/H] plane for all stars in the stellar disk of galaxy g7.55e11. We see that in the early universe self enrichment quickly raises the $\alpha$ abundance to values of [$\alpha$/Fe]$\sim0.4$. From there the high-$\alpha$ sequence evolves to higher metallicity gradually decreasing its oxygen abundance. As has been shown in figure 13 and 14 of \citet{Buck2019d} the stellar disk in these simulations forms upside-down and inside-out \citep[e.g.][]{Bird2013}. Therefore the stars in the high-$\alpha$ sequence are born hot with large vertical scale-heights.

Then, around $t\sim7$ Gyr a gas rich merger brings in fresh gas diluting the ISM to lower [Fe/H] but keeping the value of [$\alpha$/Fe] roughly constant. Successive star formation in the disk then proceeds from an ISM of comparatively lower metallicity and forms the low-$\alpha$ sequence (panel at $t\sim8$ Gyr). Importantly, the high-$\alpha$ sequence is already in place at this point in time and all newly formed stars are mostly born in the low-$\alpha$ sequence from a thin gas disk. The low-$\alpha$ sequence then evolves again to high metallicities by self enrichment. Therefore, in our simulation the youngest low-$\alpha$ stars overlap the (youngest) stars of the high-$\alpha$ sequence at the highest metallicities. 
Our simulations confirm previous notions that the $\alpha$ bimodality in the MW is formed from a merger event \citep[e.g.][]{Helmi2018,Belokurov2018} but in a completely opposite sense. In our simulations it is the low-$\alpha$ sequence which is formed by the merger and \textbf{not} the high-$\alpha$ one . Our simulations further do not need to invoke a major merger for this as all $\alpha$-bimodality creating mergers are minor mergers with virial mass ratios $\lesssim10\%$ bringing in less than $\sim10^9\Msun$ of stars. In fact, the high-$\alpha$ sequence was already established before the merger and its vertical scale height was large to begin with such that the merger did not significantly heat it further. Interestingly, such a merger leaves its imprint in the age velocity dispersion relation as a sharp increase in vertical velocity dispersion at the time of the merger as we have shown in \citet{Buck2019d} (fig. 9) and as was previously shown by e.g. \citet{Martig2014b}. Such a sharp increase in velocity dispersion is indeed observed for the MW \citep[e.g.][]{Quillen2001,Freeman2002,Minchev2018}.

\subsubsection{Chemical evolution of the ISM}
The upper panels of Fig. \ref{fig:gas_mass} show the evolution of the gas mass as a function of time within different galacto-centric radii as indicated by the line color. The lower panels show the star formation history (SFH) in annuli of width $2$ kpc centred on the radii as indicated by the colorbar. In each panel we highlight with vertical gray bars the times around which the low-$\alpha$ sequences in each of the simulations form. Comparing the gas mass evolution with these gray bars, we find that the formation of the low-$\alpha$ sequences coincide with spikes in the gas mass, indicating gas accretion events. For most of these events the spikes in gas mass are found at each radius even in the very center of the galaxies. At the bottom of each of the panels we indicate with open black dots the times at which gas rich satellites containing more than $5\%$ of the host's gas mass at that time are accreted onto the main galaxy. We define the accretion time as the point where the satellites come closer than $75$ kpc to the host. The size of the black dots indicates the fraction of gas mass $M_{\rm gas}^{\rm sat}/M_{\rm gas}^{\rm gal}(t)$ brought in with that merger event. Correlating the merger events with the formation of the low-$\alpha$ sequence and the spikes in gas mass in the ISM evolution clearly shows that gas rich mergers are causing a short increase in the ISM gas mass at each radius, even in the very centers of these galaxies.

Looking at the SFH in the lower panels we find that the galaxies with a distinct low- and high-$\alpha$ disk show a high star formation rate in the central regions ($R<5$ kpc) at early times but not particularly associated with a starburst or merger event. This is not unexpected as we have already seen that the high-$\alpha$ sequence is exclusively old (Fig. \ref{fig:age}) and mainly originating from the inner disk (Fig. \ref{fig:birth_r}). The SFHs further confirm the implications of Fig. \ref{fig:time} where the high-$\alpha$ sequence forms at early times out of quickly self-enriched gas extending to higher metallicities as time progresses while the low-$\alpha$ disk forms later after gas rich merger brought in unenriched gas and trigger star formation in the outer disk ($R>5$ kpc).

The effect of these merger events on the ISM metallicity and the [$\alpha$/Fe] ratio is investigated in figure \ref{fig:ISM}. The upper panels show the evolution of [Fe/H] as a function of time for galacto-centric annuli of width $\Delta R=2$ kpc while the lower panels show the time evolution of [$\alpha$/Fe]. Again, we indicate the time when the low-$\alpha$ sequence is formed with gray vertical bars and the times of gas rich mergers with open black dots.

In general, the coloured lines show the expected chemical evolution of the ISM of MW like galaxies \citep[e.g.][]{Minchev2018} where the iron abundance increases quickly at early times ($\lesssim4$ Gyr) with a gradual flattening of the metal enrichment at late cosmic times. The ratio of [$\alpha$/Fe] on the other hand quickly decreases from its rapidly reached high values around [$\alpha$/Fe]$\sim0.5$ at early cosmic times. For both [Fe/H] and [$\alpha$/Fe] we see that early on the lines for different galactic radii were closer together in abundance space and then the differences between average abundance ratios at each radius are increasing at late times. This might indicate that strong radial gradients were only established at later cosmic times ($>5$ Gyr) in these simulations. Again, if we correlate gray bars, black dots and the evolution of the coloured lines we find that at each merger event when the stars of lower-$\alpha$ sequence in these simulations were born the abundance ratio of [Fe/H] dropped at each radius even in the very center (blue lines). The abundance ratio of [$\alpha$/Fe] on the other is not strongly affected by the event. This confirms the hypotheses that the accretion of gas rich satellites dilutes the ISM metallicity to lower [Fe/H] values while keeping the [$\alpha$/Fe] ratio almost constant. In this way the generation of stars formed after the accretion events has lower metallicity but the same [$\alpha$/Fe] ratio and thus populates the space of the lower-$\alpha$ sequence. The scenario for the formation of the chemical bimodality of disk stars presented here is conceptually similar to the one presented in \citet{Haywood2019} except for the fact that it is not the bar which drives radial gas flows in these simulations but the accretion of gas rich satellites. In this sense it resembles the two infall scenario \citep[e.g.][]{Chiappini1997,Spitoni2019} and the exact timing and gas mass brought in by the merger determines the pattern shape. This is also why we find a variety of different patterns in \ref{fig:nihao}.

\section{Conclusion}
\label{sec:conclusion}

In this work we studied the structure of galactic stellar disks in chemical abundance space using four high-resolution simulations from the NIHAO-UHD project \citep{Buck2019d}. 
In particular, we focussed on the formation scenario of the high- and low-$\alpha$ sequences in [Fe/H] vs. [$\alpha$/Fe] space identifying the accretion events of gas rich satellites onto the main galaxy as the cause of the low-$\alpha$ sequence. Thus, the results of this study have strong implications for observationally constraining the merger history of local galaxies such as MW or Andromeda.

Our results are summarized as follows:
\begin{itemize}
\item The stellar disks of the NIHAO-UHD simulations produce a bimodal chemical abundance pattern in the [Fe/H] vs. [$\alpha$/Fe] plane  (e.g fig. \ref{fig:impression}). The kinematic, structural and spatial properties of the bimodal $\alpha$-sequence in our simulations reproduces that of observations. In all simulations, the high-$\alpha$ sequence is old and radially concentrated toward the center of the Galaxy, unlike the low-$\alpha$ sequence which is younger and more radially extended (fig. \ref{fig:impression}). Furthermore, the low-$\alpha$ sequence is concentrated to the disk mid-plane while the high-$\alpha$ stars extend to larger heights from the plane.

\item Investigating the ages of stars in the two sequences we find to first order that [$\alpha$/Fe] is a good proxy for stellar age (middle panel of fig. \ref{fig:hayden}). Looking at the two sequences we find an almost entirely old high-$\alpha$ sequence and a younger low-$\alpha$ sequence (compare fig \ref{fig:age}). Our results for the low-$\alpha$ sequence are in good agreement with observations from e.g. \citet{Silva2018} while they are at odds for the high-$\alpha$ sequence where there are observed young stars which our simulations do not predict. This might either point towards miss-labeled ages in the observations or help to improve employed simulation prescriptions. 

\item In more detail, the [Fe/H] vs. [$\alpha$/Fe] plane shows a continuous age gradient along both axes, the [$\alpha$/Fe] and the [Fe/H] axis (middle panel of fig. \ref{fig:insitu}). Our simulations predict a large age gradient at fixed [Fe/H] across all [$\alpha$/Fe]. For the low-$\alpha$ sequence we find further a slight age gradient at fixed [$\alpha$/Fe] across [Fe/H] in good agreement with results from \citet{Ness2019}. 

\item Tracing the birth radii of the stars we find that at a given present-day galacto-centric radius the high-$\alpha$ sequence consists of stars mostly from the central regions of the galaxy ($R\lesssim6$ kpc) while the low-$\alpha$ sequence is made up of stars with vastly different birth radii with contributions from both, the inner and the outer disk (Fig. \ref{fig:birth_r}). This shows that radial migration plays a major role in shaping the high- and low-$\alpha$ sequence in these simulations. 

\item The origin of the chemical bimodality of disk stars in the NIHAO-UHD simulations is a generic consequence of a gas-rich merger at some time in the galaxy's evolution similar to results obtained by \citet{Grand2018}. The high-$\alpha$ sequence is formed first in the early galaxies with a large vertical scale height already at birth (e.g. fig \ref{fig:time}). Then later, after a gas rich merger of virial mass ratio $\lesssim10\%$ ($M_{\rm star}\lesssim10^9\Msun$) brings in fresh metal-poor gas diluting the interstellar medium's metallicity the low-$\alpha$ sequence forms (fig. \ref{fig:gas_mass} and fig. \ref{fig:ISM}). Thus, our simulations confirm the idea that the chemical bimodality in the MW is caused by a merger but with completely opposite interpretations. It is the low-$\alpha$ sequence which is formed by the merger, \textbf{not} the high-$\alpha$ sequence.

\end{itemize}

The merger origin of the chemical bimodality combined with the prominent influence of radial migration has strong implications for observationally constraining the merger history of the MW or external galaxies like Andromeda.While stellar ages and birth radii are no direct observables of stars, recent observational progress enabled to infer these properties robustly. Age dating the stars in the [Fe/H]-[$\alpha$/Fe] plane can constrain the time of formation of the low-$\alpha$ sequence and might constrain the merger history of our MW.

\section*{acknowledgments}
TB thanks the anonymous referee for their careful reading of the manuscript and constructive feedback which helped to improve the quality of the paper.
A special thanks goes to Melissa Ness for carefully reading this draft and providing extensive feedback which improved the quality of this manuscript a lot! TB thanks Rob Grand, Aura Obreja, Christoph Pfrommer, Andrea V. Macci\`o, Wilma Trick and Ivan Minchev for insightful discussions and Sven Buder, Hans-Walter Rix and David Hogg for fruitful feedback on an early version of this draft. 
TB acknowledges support from the European Research Council under ERC-CoG grant CRAGSMAN-646955. TB gratefully acknowledges the Gauss Centre for Supercomputing e.V. (www.gauss-centre.eu) for funding this project by providing computing time on the GCS Supercomputer SuperMUC at Leibniz Supercomputing Centre (www.lrz.de).
This research made use of the {\sc{pynbody}} \citet{pynbody} and {\sc{tangos}} \citet{tangos} package to analyze the simulations and used the {\sc{python}} package {\sc{matplotlib}} \citep{matplotlib} to display all figures in this work. Data analysis for this work made intensive use of the {\sc{python}} library {\sc{SciPy}} \citep{scipy}, in particular {\sc{NumPy and IPython}} \citep{numpy,ipython}.
This research was carried out on the High Performance Computing resources at New York University Abu Dhabi; Simulations have been performed on the ISAAC cluster of the Max-Planck-Institut f\"ur Astronomie at the Rechenzentrum in Garching and the DRACO cluster at the Rechenzentrum in Garching. We greatly appreciate the contributions of all these computing allocations.
\bibliography{astro-ph.bib}

\appendix

\section{$\alpha$-bimodality in the lower resolution simulations}
\begin{figure*}
\begin{center}
\includegraphics[width=\textwidth]{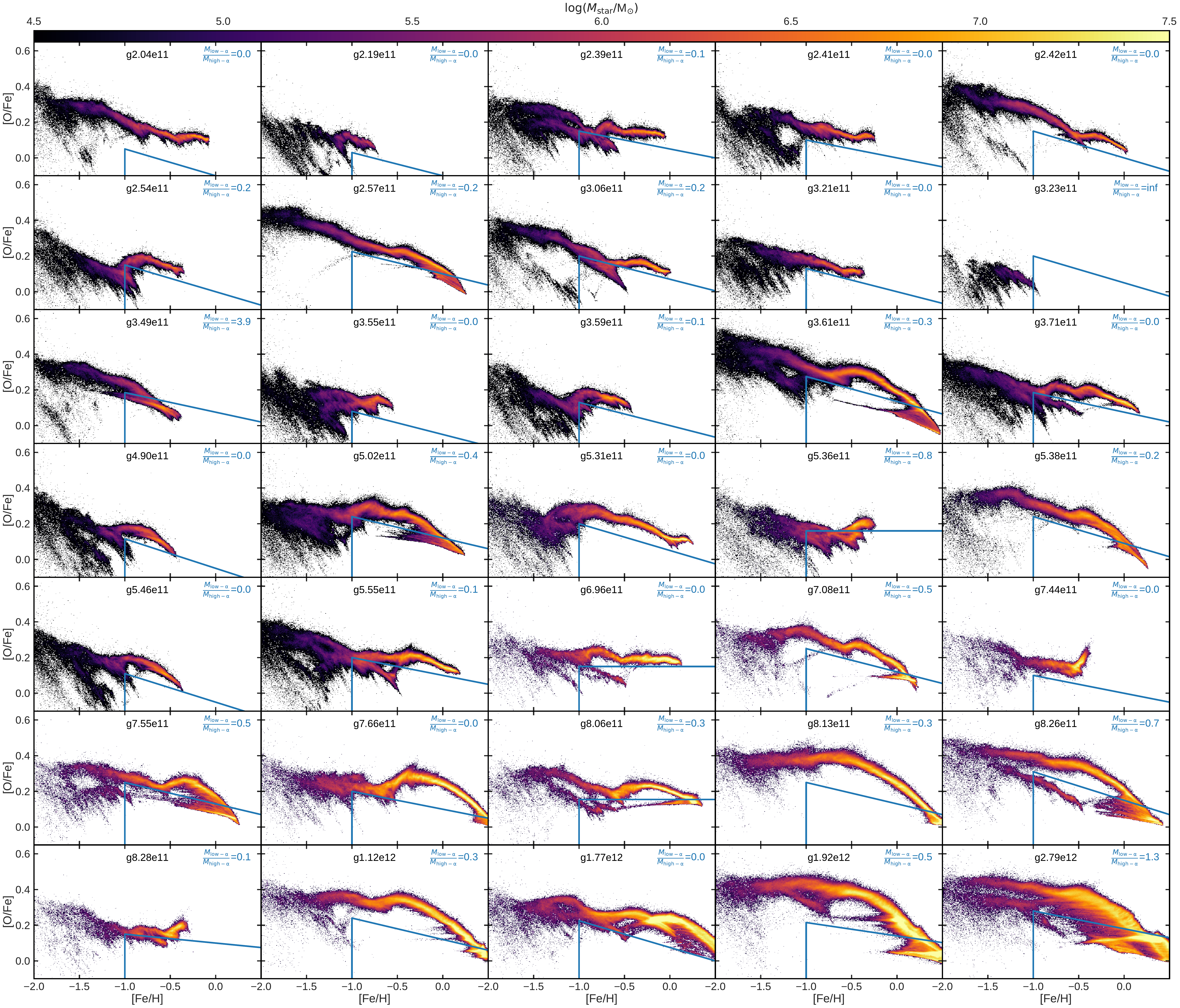}
\end{center}
\vspace{-.35cm}
\caption{[$\alpha$/Fe] vs. [Fe/H] plane for all disk stars ($0<R<20$ kpc and $-5<z<5$ kpc) of galaxies in the mass range $2\times 10^{11}<M_{\rm halo}<3\times10^{12}\Msun$ taken from the classic NIHAO simulations. Many galaxies show a bimodality in [$\alpha$/Fe] vs. [Fe/H] space, especially at larger halo masses.}
\label{fig:nihao}
\end{figure*}

Figure \ref{fig:nihao} shows the [$\alpha$/Fe] vs. [Fe/H] plane for 35 galaxies in the halo mass range $2\times 10^{11}<M_{\rm halo}<3\times10^{12}\Msun$ selected from the original NIHAO sample. We plot all disk stars ($0<R<20$ kpc and $-5<z<5$ kpc) and color code the amount of stellar mass falling into in each bin in the [$\alpha$/Fe] vs. [Fe/H] plane. The original NIHAO galaxies have roughly a 8 times lower mass resolution compared to our fiducial galaxies presented in the main text. We find a diversity in the $\alpha$ sequences ranging from single tracks (e.g. g2.57e11, g7.66e11 or g8.13e11) over a bimodality at low metallicities (e.g. 4.90e11, g5.36e11, g5.55e11, 6.96e11) to MW-like bimodalities spanning large ranges in metallicity (e.g. g3.61e11, g8.06e11, g8.26e11) and in some galaxies we even find more than two tracks (e.g. g1.92e12, g2.79e12). We quantify the strength of the bimodalities by the stellar mass ratio of the two sequences which we quote in the upper right corner of each panel. Separation of the two sequences is done by eye and we define a value larger than 0.2 as the incidence of a bimodality. Unfortunately, the stellar mass ratio of low- and high-$\alpha$ sequence in the MW is not known and will depend on the specific selection function of the survey or galactic region \citep[see e.g. Fig. \ref{fig:hayden} and e.g.][for the MW]{Hayden2015}. Thus, every close comparison to the MW bimodality would involve modelling the specific selection function of the survey used for comparison. This is outside the scope of this paper which is why we here only do a qualitative comparison.

\section{Stellar ages and birth radii in the [Fe/H] vs. [$\alpha$/Fe] plane}

\begin{figure*}
\begin{center}
\includegraphics[width=\textwidth]{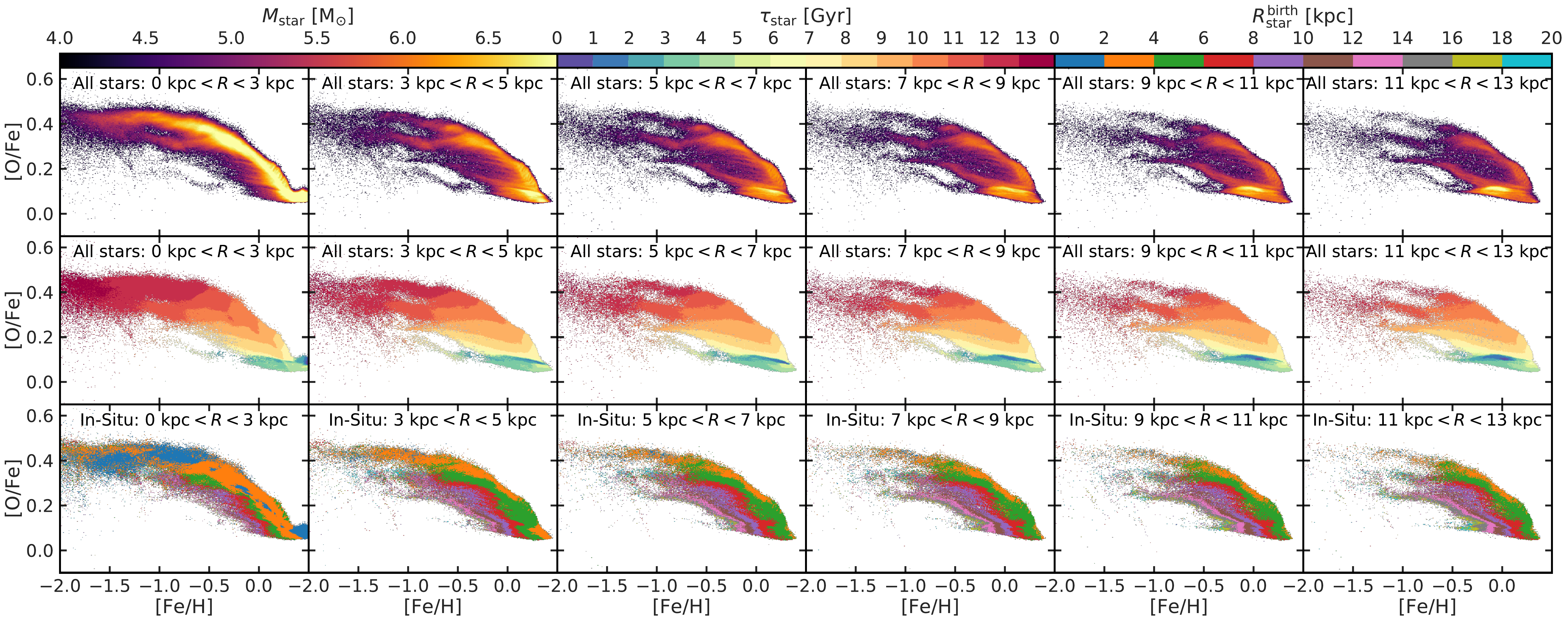}
\includegraphics[width=\textwidth]{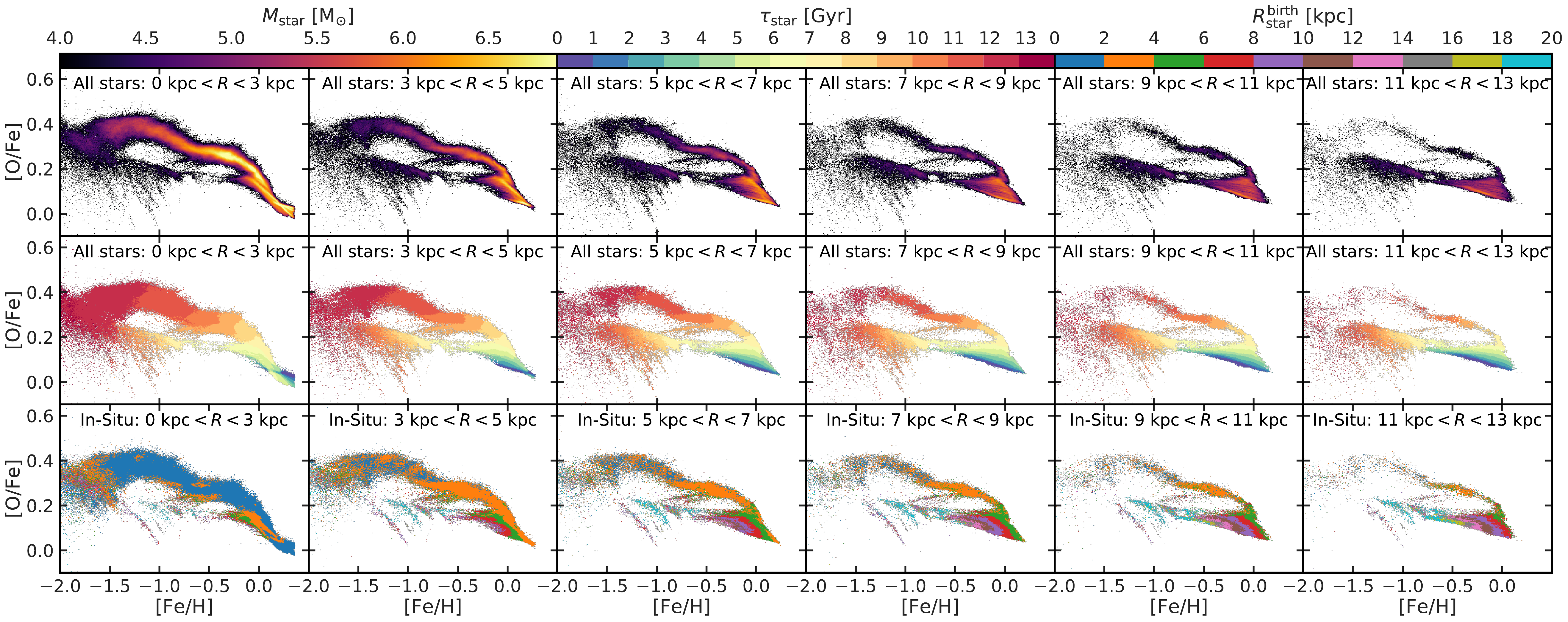}
\includegraphics[width=\textwidth]{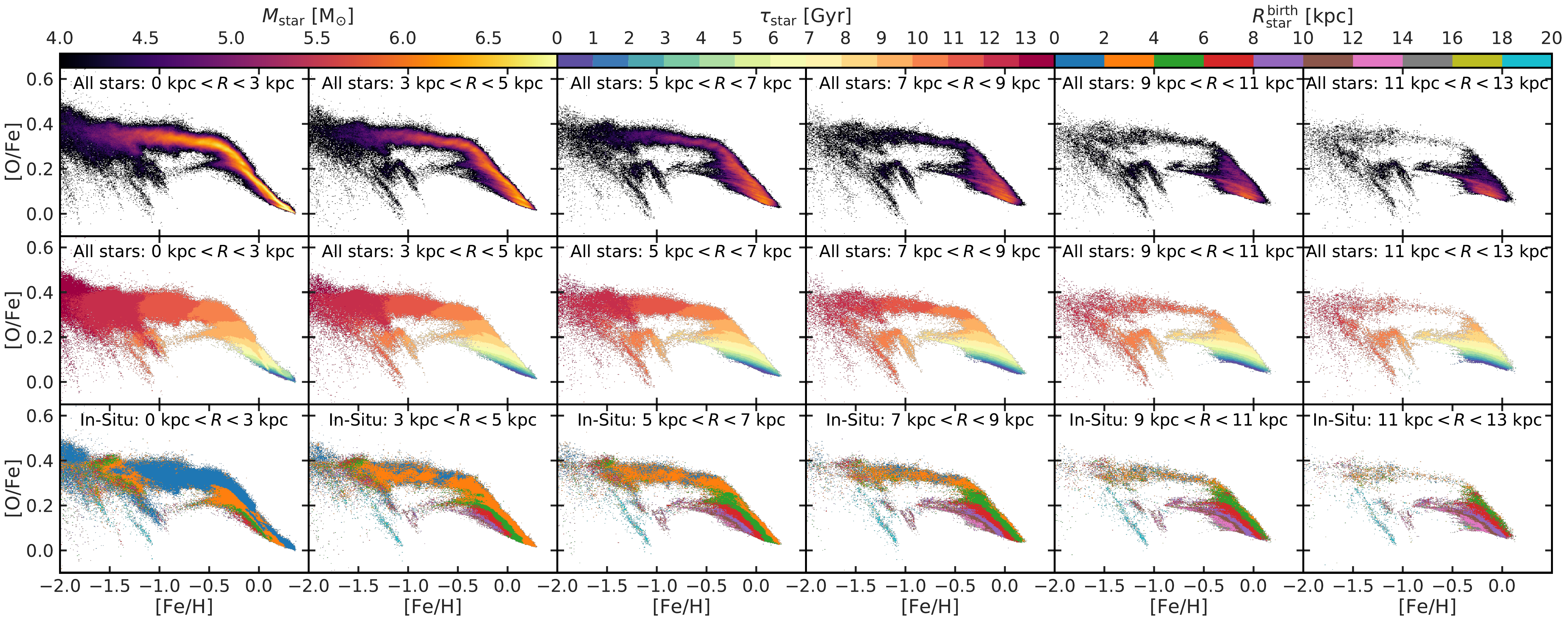}
\end{center}
\vspace{-.35cm}
\caption{Same as figure \ref{fig:insitu} but this time showing galaxy g2.79e12 (top panels), g7.55e11 (middle panels) and g7.08e11 (bottom panels). From left to right we show the different radii already presented in figure \ref{fig:impression}. This figure shows that the results drawn from figure \ref{fig:insitu} only using the star particles in the solar circle are robust across different galacto-centric radii.}
\label{fig:app1}
\end{figure*}

\begin{figure*}
\begin{center}
\includegraphics[width=\textwidth]{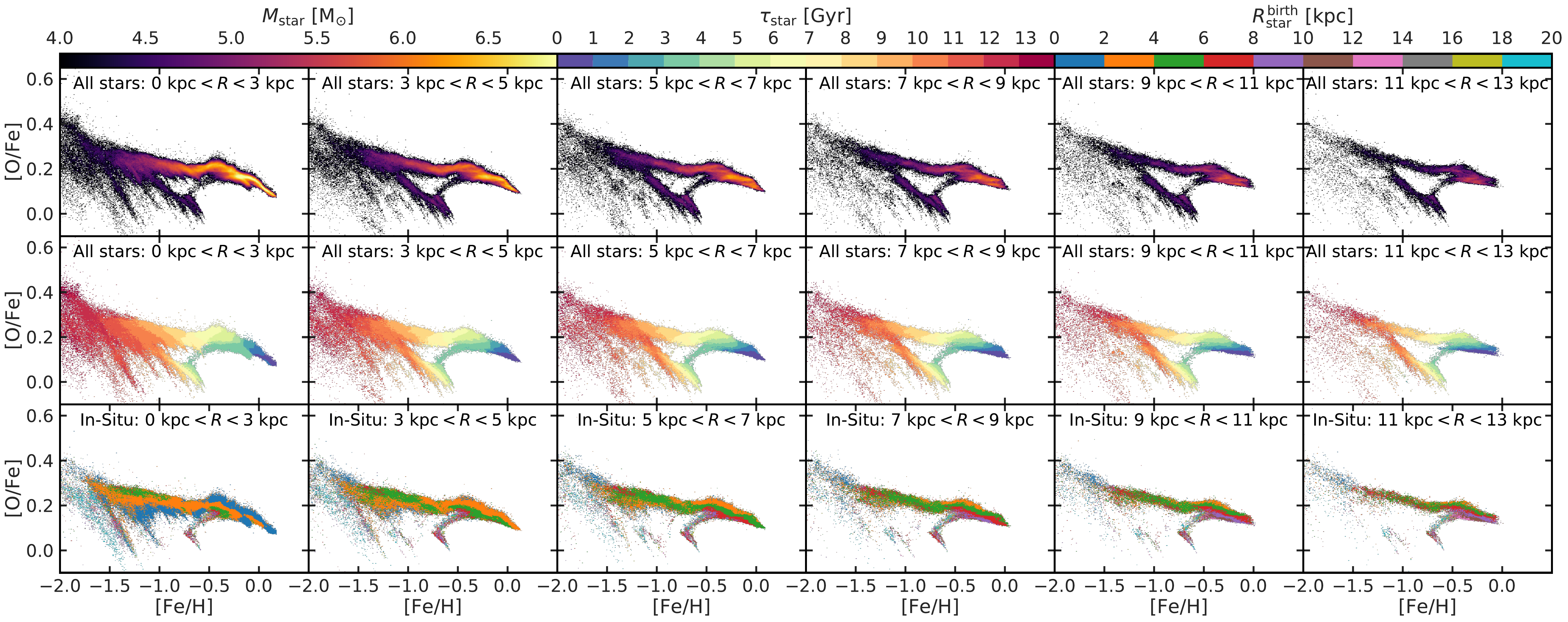}
\end{center}
\vspace{-.35cm}
\caption{Same as figure \ref{fig:insitu} but this time showing galaxy g6.96e11. From left to right we show the different radii already presented in figure \ref{fig:impression}. This figure shows that the results drawn from figure \ref{fig:insitu} only using the star particles in the solar circle are robust across different galacto-centric radii.}
\label{fig:app2}
\end{figure*}

Figure \ref{fig:app1} and \ref{fig:app2} show a variant of figure \ref{fig:insitu} for all galcto-centric annuli. The difference between figure \ref{fig:insitu} and thgese figures is that here we show in the first two rows all stars in the stellar disk at present-day while in the bottom panel we only show stars which were born inside the stellar disk excluding accreted stars as indicated in the panels. The color coding in the single panels is the same as in figure \ref{fig:insitu}. Comparing the upper and lower panels shows that accreted stars are mostly found at low metallicities. Comparing on the other hand the panel at $8$ kpc with the panels for the other radii bins we find that the results drawn from figure \ref{fig:insitu} are robust across different galacto-centric radii.
 
\label{lastpage}
\end{document}